\documentclass[12pt]{article}

\usepackage{putex}

\usepackage{cancel}
\usepackage{simplewick}
\usepackage{multirow}
\usepackage{comment}

\usepackage{graphicx}
\usepackage{caption}
\usepackage{amsmath}
\usepackage{array}
\usepackage{subcaption}
\usepackage{epstopdf}
\usepackage{enumerate}
\usepackage{cite}
\usepackage{youngtab}
\usepackage{tensor}
\usepackage{slashed}
\usepackage[aligntableaux=center]{ytableau}
\usepackage[utf8]{inputenc}
\usepackage{rotating}
\usepackage{bigfoot}

\usepackage[
      colorlinks=true,
      linkcolor=black,
      urlcolor=blue,
      filecolor=black,
      citecolor=red,
      ]{hyperref}

\usepackage[compat=1.1.0]{tikz-feynman}
\usetikzlibrary{decorations.markings}

\numberwithin{equation}{section}

\def \( {\left(}
\def \) {\right)}
\def \< {\left<}
\def \> {\right>}

\newcommand{\be}{\begin{equation}} \newcommand{\ee}{\end{equation}}
\newcommand{\bea}{\begin{eqnarray}}  \newcommand{\eea}{\end{eqnarray}}
\newcommand{\nn}{\nonumber}

\tikzfeynmanset{
	O/.style = {very thick}
}
\tikzfeynmanset{
	Ot/.style = {scalar}
}
\tikzfeynmanset{
	Th/.style = {double distance=2pt}
}
\tikzfeynmanset{
	dO/.style = {fermion}
}

\begin{document}

\vspace*{-1.5cm}
\begin{flushright}    
  {\small
  }
\end{flushright}

\vspace{1.8cm}
\begin{center}        
  \Huge On Convexity of Charged Operators in CFTs \\
and the Weak Gravity Conjecture
\end{center}

\vspace{0.7cm}
\begin{center}        
{\large  Ofer Aharony$^1$ and Eran Palti$^{2}$}
\end{center}

\vspace{0.15cm}
\begin{center}        
\emph{$^1$ Department of Particle Physics and Astrophysics, Weizmann Institute of Science, \\ Rehovot 7610001, Israel}\\[.3cm]
\emph{$^2$ Department of Physics, Ben-Gurion University of the Negev, Beer-Sheva 84105, Israel}\\[.4cm]
             
e-mail: \tt ofer.aharony@weizmann.ac.il, palti@bgu.ac.il
\end{center}

\vspace{1.5cm}


\begin{abstract}
\noindent The Weak Gravity Conjecture is typically stated as a bound on the mass-to-charge ratio of a particle in the theory. Alternatively, it has been proposed that its natural formulation is in terms of the existence of a particle which is self-repulsive under all long-range forces. We propose a closely related, but distinct, formulation, which is that it should correspond to a particle with non-negative self-binding energy. This formulation is particularly interesting in anti-de Sitter space, because it has a simple conformal field theory (CFT) dual formulation: let $\Delta(q)$ be the dimension of the lowest-dimension operator with charge $q$ under some global $U(1)$ symmetry, then $\Delta(q)$ must be a convex function of $q$. This formulation avoids any reference to holographic dual forces or even to locality in spacetime, and so we make a wild leap, and conjecture that such convexity of the spectrum of charges holds for any (unitary) conformal field theory, not just those that have weakly coupled and weakly curved duals. This Charge Convexity Conjecture, and its natural generalization to larger global symmetry groups, can be tested in various examples where anomalous dimensions can be computed, by perturbation theory, $1/N$ expansions and semi-classical methods. In all examples that we tested we find that the conjecture holds. We do not yet understand from the CFT point of view why this is true.
\end{abstract}

\thispagestyle{empty}
\clearpage

\tableofcontents

\setcounter{page}{1}



\section{Introduction}
\label{sec:intro}

In this note we consider properties of local operators in unitary Conformal Field Theories (CFTs) with continuous global symmetries (in $d > 2$ space-time dimensions). We propose that such operators should satisfy a certain convexity-like property:

{\bf Abelian Convex Charge Conjecture:\;}{\it Consider any CFT with a $U(1)$ global symmetry. Denote by $\Delta\left(q\right)$ the dimension of the lowest dimension operator of charge $q$. Then this must satisfy a convex-like constraint
\be
\Delta\left(n_1 q_0+n_2q_0\right) \geq \Delta\left(n_1 q_0\right) + \Delta\left(n_2 q_0\right) \;,
\label{ccc}
\ee
for any positive integers $n_1$, $n_2$, for some $q_0$ of order one.}

Equivalently, the conjecture states that the operator product expansion (OPE) of the lowest-dimension operators with positive charges has no singular terms, whenever the charges are integer multiples of $q_0$. The $U(1)$ charge can always be normalized to be an integer (with the minimal charge equal to one). As we will discuss below, the non-trivial statement here is that $q_0$ is not parameterically large in any parameter of the CFT. We propose that this is an exact property of any CFT whose continuous global symmetry is precisely $U(1)$.

There is a natural generalization of the proposal also to theories with larger symmetry groups :

{\bf Convex Charge Conjecture:\;}{\it Consider any CFT with a continuous global symmetry group $G$, and consider a simple factor $G_0 \subset G$. Denote by $\Delta\left(r\right)$ the dimension of the lowest dimension operator in the representation $r$ of $G$. Then, there is always some representation $r_0$, which is non-trivial in $G_0$ and has weights of order one, such that the dimensions ${\tilde \Delta}(q) \equiv \Delta({\rm Sym}^q(r_0))$ satisfy a convex-like constraint
\be
{\tilde \Delta}\left(n_1+n_2\right) \geq {\tilde \Delta}\left(n_1\right) + {\tilde \Delta}\left(n_2\right) \;,
\label{cccg}
\ee
for any positive integers $n_1$, $n_2$.}\footnote{By ${\rm Sym}^q(r_0)$ we mean the symmetric product of $q$ copies of the representation $r_0$. In cases where this symmetric product is reducible, we mean the specific representation whose highest weight is $q$ times the highest weight of $r_0$.}

Our proposal is motivated by, but is much more general than, the Weak Gravity Conjecture \cite{Arkani-Hamed:2006emk} (WGC) in Anti-de Sitter space. Indeed, we propose that the WGC should be formulated (also in flat space) as a statement about the self-binding energy of a particle:

{\bf Positive Binding Conjecture:} {\it For a (weakly coupled) gravitational theory with a $U(1)$ gauge field, there should exist at least one charged particle in the theory, with charge of order one, which has a non-negative self-binding energy.} 

By a self-binding energy here we mean the difference of energies between the lowest two-particle state and twice the energy of the one-particle state. This formulation is closely related to previous formulations, most closely to self-repulsive statements \cite{Palti:2017elp,Heidenreich:2019zkl}, but has some important differences which become more pronounced in anti-de Sitter (AdS) space. Specifically, in AdS space the binding energy receives important contributions from contact terms, and not only from long-range forces.

The organization of this paper is as follows. In section \ref{sec:wgcbe} we discuss the formulation of the WGC in terms of binding energy, in flat space and in anti-de-Sitter space. In section \ref{sec:wgcad} we introduce the CFT dual statements, in terms of convexity of the spectrum, and discuss some related details. In section \ref{sec:gencft} we discuss how the conjecture fits in with what is known generally about CFTs. Our discussion in these sections is mostly for the $U(1)$ case, but it can be generalized in a straightforward manner also to the non-Abelian cases mentioned above (at least in dimensions where the gauge theory in the bulk is IR-free).
In section \ref{sec:tests} we perform preliminary tests of the conjecture in many different simple examples of CFTs for which the anomalous dimensions are computable in some expansion, and we find that it is always satisfied. We also find that it is not satisfied in theories like the $O(N)$ model in $4<d<6$, which are believed not to be unitary. We end with a summary in section \ref{sec:summary}.

\section{The Weak Gravity Conjecture and Binding Energy}
\label{sec:wgcbe}

The Swampland program aims to understand constraints on effective theories coming from the requirement of an ultraviolet completion into quantum gravity \cite{Vafa:2005ui} (see \cite{Palti:2019pca,vanBeest:2021lhn} for reviews). One such conjectured constraint is the Weak Gravity Conjecture \cite{Arkani-Hamed:2006emk}, which places a bound on the mass of a charged particle under a $U(1)$ gauge field -- there must be a particle obeying
\be
m \leq \sqrt{2} g q M_p \;.
\label{EWGC}
\ee
Here $m$ is the mass of the particle, $g$ is the gauge coupling, $q$ its quantised charge, and $M_p$ the Planck scale. There are no known counter-examples to the conjecture coming from string theory to date. Here we wrote down the equation for the case of a four dimensional space-time, but it can be generalized to other dimensions as well.

In \cite{Palti:2017elp} (see also \cite{Heidenreich:2019zkl}) it was argued that the more precise formulation of the WGC should be in terms of the existence of a self-repulsive charged particle. This formulation differs from (\ref{EWGC}) in the presence of massless scalar fields, since they mediate an additional self-attractive long-range force. Specifically, the generalization to this case reads
\be
2 g^2 q^2 M_p^2 \geq m^2 + \mu^2 M_p^2 \;,
\label{repsca}
\ee
where $\mu$ is the coupling of the particle to the scalar fields. This has been tested in string theory, for example see \cite{Lee:2018urn,Lee:2018spm,Heidenreich:2019zkl}. 

One motivation for the self-repulsive statement is that it ensures the absence of a large number $n \sim g^{-1}$ of stable bound states formed from copies of the particle with the largest charge-to-mass ratio in the theory \cite{Arkani-Hamed:2006emk}. It is as yet unclear why such states would be inconsistent. But we can assume such a requirement and study its consequences. 

The reason that a self-attractive charged particle would form a stable bound state with itself (or $n$ copies of itself) is because the charge of the bound state would be twice that of the particle, while the energy of the bound state would be less than twice the mass of the particle due to the binding energy. If that particle is the one with the largest charge-to-mass ratio in the theory, then the bound state cannot decay to it or to any other state in the theory. 

If one is motivated by the absence of such stable bound states, then actually the most direct and natural formulation of the WGC is as the Positive Binding Conjecture discussed in the introduction. 

In flat space we can apply this statement to a particle placed an infinite distance from its copy. In that case, the binding energy positivity is mapped directly to the particle being self-repulsive under the long-range Coulomb forces\footnote{More precisely, this is true for four non-compact space-time dimensions. In higher dimensions it is possible \cite{Heidenreich:2019zkl} that the long-distance force could be attractive but still there will not be any bound state.}, yielding the formulation (\ref{repsca}). However, global AdS space behaves effectively like a box, so there is a limit as to how far the particle can be displaced from its copy. In particular there is no obvious relation between the long-range force and the binding energy. This makes the binding energy formulation of the WGC somewhat different from both the statements about unstable Reissner-Nordstrom black holes (\ref{EWGC}) and about repulsive long-range forces (\ref{repsca}). In AdS space there is a contribution from contact interactions to the binding energy, which cannot be dismissed. 

Let us make this more precise for the case of 5-dimensional AdS space, following the calculation in \cite{Fitzpatrick:2011hh}. We consider a 5-dimensional gravitational theory with a $U(1)$ gauge field and a scalar field $\varphi$ charged under it with charge $q$, that has an $AdS_5$ solution:
\bea
S &=& \frac{1}{\kappa^2} \int d^5x \sqrt{-g} \left(\frac12 R + 6 - \frac{1}{4g^2} F_{\mu \nu}^2 + \left|D_{\mu} \varphi \right|^2  \right. \nn \\
& & \left. \qquad\qquad\qquad - m^2 \left|\varphi\right|^2 - a \left|\varphi\right|^4 + b \left|\varphi\right|^2 \left|D_{\mu} \varphi \right|^2 \right)  \;.
\label{5dadsgr}
\eea
Here $\kappa$ is the 5-dimensional Planck scale. We measure every dimensionful quantity in units of the AdS radius $R_{AdS}$, and set this radius to unity $R_{AdS}=1$. The coefficients $a$ and $b$ are arbitrary constants. The action (\ref{5dadsgr}) captures all the relevant contributions to the self-binding energy of $\varphi$ (up to two-derivative order). Let us denote the self-binding energy of $\varphi$ as $\gamma_{\varphi^2}$. This was calculated at leading order in perturbation theory in \cite{Fitzpatrick:2011hh}, which found
\be
\gamma_{\varphi^2} = \gamma_{\varphi^2}^{\mathrm{photon}} + \gamma_{\varphi^2}^{\mathrm{graviton}} + \gamma_{\varphi^2}^{\mathrm{quartic}} \;,
\ee
with
\bea
\gamma_{\varphi^2}^{\mathrm{photon}} &=& \frac{\pi^2 N_{\Delta}^4 g^2 q^2}{2\Delta-1}\;, \nn \\
\gamma_{\varphi^2}^{\mathrm{graviton}} &=& -\frac{2\pi^2 N_{\Delta}^4 \Delta^2\left(\Delta-2\right)}{3\left(\Delta-1\right)\left(2\Delta-1\right)}\;, \nn \\
\gamma_{\varphi^2}^{\mathrm{quartic}} &=& \frac{\pi^2 N_{\Delta}^4 \left(a + b\Delta\left(2-\Delta\right)\right)}{\left(\Delta-1\right)\left(2\Delta-1\right)}\;.
\eea
Here we have introduced $\Delta$ (the dimension of the CFT operator dual to $\varphi$) and $N_{\Delta}$ as
\be
m^2 = \Delta \left( \Delta - 4 \right) \;,\;\; N_{\Delta} = \sqrt{\frac{\Delta-1}{2\pi^2}} \;.
\ee

The Positive Binding Conjecture implies that in all UV-complete theories $\gamma_{\varphi^2} \geq 0$, and there are no counter-examples to this as far as we know. In the case where the action (\ref{5dadsgr}) is completed into a supersymmetric one, and $\varphi$ is taken as a BPS state, we have the relations
\be
\mathrm{SUSY}\;: -g^2q^2 = a + b\Delta\left(2-\Delta\right) = -\frac23 \Delta^2  \;.
\ee
This leads to the exact relation $\gamma_{\varphi^2} =0 $. It also manifests explicitly that the contact contribution to the binding energy becomes negligible relative to the other contributions in the $\Delta \rightarrow \infty$ limit, where the particle is very heavy compared to the AdS scale.

Note that the action (\ref{5dadsgr}) is not the most general low-energy effective action, in that it does not include the contribution from additional massless scalar fields. But the main point is that in AdS the contact interactions are important, and so the binding energy formulation of the WGC is significantly different from the one based on black holes or long-range forces. It is also the one which is precisely saturated with supersymmetry, and so it seems like the more natural one.


\section{A Convex Dimension Conjecture}
\label{sec:wgcad}

The AdS/CFT correspondence \cite{Maldacena:1997re,Witten:1998qj,Gubser:1998bc} maps gravitational theories on $(d+1)$-dimensional AdS space to $d$-dimensional CFTs. This correspondence maps the energy of a state in global AdS space to the dimension of the corresponding local operator (related to the state by the state/operator correspondence) in the CFT, and it maps gauge fields in the bulk to global symmetries of the CFT.\footnote{The last statement depends on the boundary conditions, but it is always true for $d \geq 3$. In any case the inverse statement is always true; a continuous global symmetry in the CFT is related to a gauge field in the bulk.} Thus, it is natural to try to map the WGC in AdS space to a statement about CFTs, which should hold at least for CFTs with weakly coupled and weakly curved gravitational duals, and, as we propose, also more generally.

The holographic dual of the WGC in AdS was first explored in \cite{Nakayama:2015hga}, where it was formulated as a statement on the ratio of the dimension to the charge of an operator in the CFT (for 4-dimensional CFTs)
\be
\frac{\Delta^2}{q^2} \leq \frac{9}{40} \frac{C_T}{C_V} \;.
\ee
Here $\Delta$ is the dimension of the operator, $q$ its charge, and $C_T$ and $C_V$ are the coefficients of the two-point functions of the energy-momentum tensor and the global symmetry current, respectively (schematically, these measure the number of degrees of freedom in the CFT, with $C_T$ counting all of them and $C_V$ only the ones charged under the global $U(1)$). Subsequently, the holographic dual of the Weak Gravity Conjecture was further developed in \cite{Giombi:2017mxl,Benjamin:2016fhe,Montero:2016tif,Bae:2018qym,Montero:2017mdq,Urbano:2018kax,Montero:2018fns,Agarwal:2019crm,Alday:2019qrf}. In \cite{Baume:2020dqd,Perlmutter:2020buo} the holographic dual of the distance conjecture was developed.

In this paper we discussed a different formulation of the WGC in terms of binding energy, and this has a simpler formulation in the dual CFT, as the Convex Charge Conjecture (\ref{ccc}) that we presented in the introduction.\footnote{Note that (\ref{ccc}) is not quite equivalent to convexity of the function $\Delta\left(q\right)$. More precisely, (\ref{ccc}) follows if $\Delta\left(q\right)$ is convex as a function for all real $q \geq 0$. But since in the CFT $\Delta(q)$ is defined only over the integers (and more precisely, in the conjecture it appears just for integer multiples of $q_0$), it is possible to have $\Delta\left(q\right)$ which is not convex as a real function, and still satisfy (\ref{ccc}).} While the motivation for the conjecture comes from CFTs with weakly coupled and weakly curved gravitational duals, we will boldly conjecture that it holds for all CFTs, and perform various tests of this conjecture below\footnote{Note that our conjecture is not directly related to previous conjectures on the convexity of the black hole spectrum in flat space made by N. Arkani-Hamed \cite{Nima}, where it was conjectured that $m/q$ is a convex function of the charge $q$ in flat space.}.

In the case of weakly-coupled theories, we can formulate the conjecture in terms of specific operators related to the fields appearing in the action. For example, consider a primary scalar operator $\phi$ charged under a global symmetry, with charge of order one. Denote by $\phi^n$ the first (lowest dimension) primary operator appearing in the (symmetrised) OPE of $n$ $\phi$'s. Then, the dimension of $\phi^n$ should satisfy (\ref{ccc}) (or its generalization (\ref{cccg})). We can introduce a more condensed notation as
\be
\gamma_{n_1,n_2} \equiv \Delta\left(\phi^{n_1+n_2}\right) - \Big( \Delta\left(\phi^{n_1}\right) + \Delta\left(\phi^{n_2}\right) \Big) \;,
\label{gn1n2def}
\ee
and then write  (\ref{ccc}) or (\ref{cccg}) as
\be
\gamma_{n_1,n_2} \geq 0 \;.
\label{gccc}
\ee
We will use this notation in the weakly-coupled tests that we will perform below. 

There are various extensions of the Weak Gravity Conjecture that have been studied in the literature, see the review \cite{Palti:2019pca}, which will also play a role here. Specifically, it is possible that when the global symmetry group $G$ contains multiple $U(1)$'s that can mix together, one has to define a more subtle notion of convexity in the multi-dimensional space of $U(1)$ charges, perhaps along the lines of the analysis of \cite{Montero:2016tif,Heidenreich:2016aqi,Andriolo:2018lvp,Heidenreich:2019zkl}. A possible simple restriction which avoids needing to consider this higher dimensional space is that if there are $n$ $U(1)$'s, then the conjecture should hold when $G_0$ is chosen to be one of $n$ specific linearly independent combinations of the $U(1)$'s. However, we expect that stronger statements are also likely to hold.

We note that all the examples we will discuss below are consistent with a stronger version of our conjecture, which is that the charge $q_0$ (or the representation $r_0$ which appears) is the charge of the charged operator with the lowest dimension in the CFT. This is related to stronger versions of the WGC, stating that it should hold for the charged particle with the smallest mass. Such strong versions of the Weak Gravity Conjecture have been studied on the gravity side, starting already from the original paper \cite{Arkani-Hamed:2006emk}, and we refer to the review \cite{Palti:2019pca} for more details. Note that in the case of multiple $U(1)$'s there is a candidate counter-example to such a statement proposed in \cite{Heidenreich:2016aqi}. More specifically, it is proposed as a possible counter-example to the statement that along any ray in the $U(1)$ charge space the lightest state satisfies the strong Weak Gravity Conjecture. 
In the case of a single $U(1)$ there are no proposed counter-examples in string theory to this strong version.\footnote{We note that another possible strong version of the conjecture would be that $q_0$ should be associated to the smallest possible charge, but the example in section \ref{sec:susy} contradicts this, as does the example in \cite{Heidenreich:2016aqi}.}

Another possible stronger version of the conjecture, which is consistent with all of our examples, is that the spectrum of all charged scalar operators is convex. In particular, $q_0$ can always be chosen to be the lowest charge of a charged scalar operator.

\section{Application to General CFTs}
\label{sec:gencft}

Before discussing the conjecture for general CFTs,
let us begin by reviewing the behavior of the spectrum of operators $\Delta(q)$ in general CFTs (see \cite{Gaume:2020bmp} for a recent review), at very large charge $q$ for a $U(1)$ global symmetry (where large here means larger than any other large parameter that the CFT may have). Recall that $\Delta(q)$ is the same as the energy of the lightest state of charge $q$ in the CFT on $S^{d-1}$. It is believed (though not proven) that in all CFTs this behavior falls into one of three classes:

\begin{itemize}

\item In generic CFTs, one expects the large charge behavior on a sphere to be similar locally to the behavior at large charge density in flat space. In this regime, the low-energy theory includes a Nambu-Goldstone boson related to the spontaneous breaking of the global symmetry. This leads to a universal behavior \cite{Hellerman:2015nra,Monin:2016jmo,Alvarez-Gaume:2016vff} of the energy at large $q$, $\Delta(q) = A q^{d/(d-1)} + \cdots$, where the constant $A$ depends on the specific theory (and a specific subleading term in $\Delta(q)$ is universal and can be computed explicitly \cite{Hellerman:2015nra,Cuomo:2020rgt}).

\item In free scalar theories where the scalar $\phi$ is charged under the $U(1)$, the operators $\phi^q$ give $\Delta(q) = q \Delta(1)$. A similar behavior arises also in supersymmetric theories with a moduli space, whenever there is some BPS operator ${\cal O}(x)$ carrying the global charge which can obtain an expectation value on the moduli space, since this implies $\Delta({\cal O}^n) = n \Delta({\cal O})$. The operators ${\cal O}^n$ are BPS operators in this case, for all $n$ (namely, they do not obey any relations in the chiral ring). Note that generally theories with moduli spaces have a continuous R-symmetry group, and then the charged operators discussed here would always carry some R-symmetry charge (and perhaps also additional charges).

\item In free fermionic theories, where the fermion carries $U(1)$ charge, the lightest charged operators are not just products of fermions, but rather because of Pauli's principle they involve products of fermions and their derivatives. This leads to states filling a Fermi sea (similar to the situation in flat space). In this case one also finds $\Delta(q) = A q^{d/(d-1)}$, where the constant $A$ depends on the type of spinor and on $d$, but the low-lying excitations over the Fermi sea look different than in the generic case mentioned above. It is interesting that in this case the function $\Delta(q)$ is not analytic even in the large $q$ limit beyond the leading order \cite{Komargodski:2021zzy}, but this does not affect our analysis.

\end{itemize}

Note that in all these cases, the large charge spectrum is convex (marginally so in the free scalar / SUSY case), so that for a large enough $q_0$ (where the large charge expansion in valid) our conjecture is always satisfied. In the second and third cases, we can compute the spectrum exactly and the conjecture is satisfied (for $q_0$ that is equal to the charge of the free scalar/fermion, or to the lowest charge carried by a BPS operator ${\cal O}$ that is non-zero on the moduli space). In the fermion case, if we take $q_0$ equal to the number of components of the fermion field, the spectrum is not even marginally convex \cite{Komargodski:2021zzy}. This means that convexity (for that value of $q_0$) is maintained under any small-parameter perturbation of the free fermion theory. Therefore, non-trivial tests of the conjecture in fermionic theories require a large number of fermion fields, for example as studied below in section \ref{sec:bankszaks}.

Thus, the non-trivial question is whether the conjecture is satisfied in the first case, for some $q_0$ of order one. This conjecture is only non-trivial if the CFT has some large parameter (which can be the inverse of some small parameter), since in this case the large charge behavior described above only sets in when $q$ scales as some power of this large parameter. Our conjecture is that for smaller values of $q$, even though the large charge expansion is not valid, convexity of the spectrum still holds, with $q_0$ of order one (namely, not scaling with the large parameter).


The original motivation for the conjecture, discussed above, is relevant for CFTs with a weakly-curved and weakly-coupled gravitational dual (in such theories the large charge expansion mentioned above is related to large black holes in AdS space). But since there is no sharp distinction in the space of CFTs between such CFTs and general CFTs, it is believed that any CFT has in some sense a quantum gravity AdS dual, and it seems reasonable to assume that any conjecture applying to ``holographic CFTs'' could apply to all CFTs. It is therefore natural to formulate the general version of the conjecture (\ref{ccc}) (or (\ref{cccg})), which would apply to any CFT.

We do not have any direct arguments for this conjecture, but we can check it in various examples (see below), corresponding to various types of expansions where the dimensions $\Delta(q)$ are computable, and it seems to always hold. In generic perturbative expansions in some parameter $\epsilon$ (which can be for instance a small coupling, the space-time dimension, or the inverse number of fields), the leading order correction to $\Delta(q) - q \Delta(1)$ will go as $\epsilon^a P(q)$ for some positive $a$ and for some polynomial $P(q)$ (often proportional to $q(q-1)$), whose leading term can be positive or negative. In the first case, convexity holds at the leading order in perturbation theory. Higher orders in perturbation theory cannot change this within the regime of validity of perturbation theory, so one has to go beyond perturbation theory to provide further checks of the conjecture. On the other hand, if the first correction is negative, then the spectrum is not convex up to some $q_0$ going as a negative power of $\epsilon$, contradicting the conjecture (\ref{ccc}). Thus, every example where $\Delta(q)$ is amenable to a perturbative expansion gives us a single test of the conjecture, and so far it has successfully passed all of these tests.

It would be nice to test our conjecture in the framework of CFTs with a weakly-curved and weakly-coupled gravitational dual (where perturbative computations in the bulk are valid). However, the specific examples of such theories that are under control tend to be supersymmetric, and the charged spectrum typically contains some BPS operator ${\cal O}(x)$. As mentioned above, if this operator parameterizes a moduli space, then $\Delta(q)$ behaves linearly and the conjecture is obvious. However, even if there is no moduli space, so that ${\cal O}^n$ is not a BPS operator for large enough values of $n$, we still have for all $n$ $\Delta({\cal O}^n) \geq n \Delta({\cal O})$ by the BPS bound. While this by itself does not imply the full convexity conjecture (\ref{ccc}), it does imply that the first non-BPS operator will obey it (and obviously all the BPS operators do), and in particular the leading perturbative correction will be consistent with the conjecture (assuming that the charge of ${\cal O}$ is not parameterically large). Thus one has to go beyond perturbation theory to check the full conjecture in these cases.

Note that in theories with such bulk duals, the spectrum naively contains also small extremal black holes (with a size much smaller than the AdS radius). There are then two options. In many cases (and in particular in supersymmetric examples) the extremal black holes are BPS (and saturate the weak gravity conjecture in the flat space limit); in this case the small extremal black holes and the multi-particle BPS states all have a linear spectrum $\Delta(q)$ with the same slope. The other option allowed by the weak gravity conjecture in flat space is that the extremal black holes may have a lower charge-to-mass ratio than some light charged particle. In this case if we consider the extremal black hole states, they would not obey the convexity condition (\ref{ccc}). However, precisely in this case the extremal black holes are also unstable towards decay to light particles, so the lowest-energy states with a specific charge would have energies much smaller than the small extremal black holes, avoiding a contradiction with the conjecture.

Note also that our conjecture about convexity of the spectrum as a function of the charge is not related as far as we can see to the convexity of the spectrum as a function of the spin (at least for large spin), which was analyzed in \cite{Komargodski:2012ek}. In particular, in the case of spin, the twist spectrum goes to a constant in the large spin limit, while in the case of charge, in generic theories $\Delta(q)$ grows faster than linearly at large $q$. It would be interesting to investigate the spectrum as a function of both the charge and the spin, but this is beyond the scope of the current paper (the results of \cite{Li:2015rfa} should be useful for this).

We have restricted our conjecture to $d>2$, so that it does not apply to two-dimensional CFTs. It is possible to find CFTs in two dimensions which have an integer parameter $N$, such that for large $N$, $q_0$ needs to be taken to scale like $N$.\footnote{We thank C. Vafa for bringing these examples to our attention, and M. Montero and G. Shiu for discussions on relevant work in progress on similar themes.} Note that for two-dimensional CFTs the dual three dimensional bulk physics is very different from higher dimensions, since gravity is not dynamical and gauge fields can be massive.

\section{Tests in Specific Theories}
\label{sec:tests}

In this section we discuss tests of our conjectures (\ref{ccc}) and (\ref{cccg}) in specific examples.
As mentioned above, from the gravitational side the conjecture does not offer significantly new testing opportunities which go beyond existing tests of the WGC. The non-trivial new tests that can be performed are from the CFT side, in theories which do not have weakly-curved gravitational duals. This also shows the power of the binding energy formulation, since from the gravity side such theories are typically not local, and there are no notions of forces and distance. Further, the gravitational and $U(1)$ gauge fields in the bulk are part of an infinite tower of higher spin fields (which are all light for weakly coupled CFTs), and so are not particularly distinguished as mediators of interactions between the states. The binding energy automatically sums over all interactions, including higher-spin modes.

Note that when the lightest charged operator is a scalar, the spectrum in the free theory is linear in the charge, so the first perturbative correction has to take a specific sign for the perturbative spectrum to be convex. On the other hand, when the light charged operators are fermions, the spectrum is convex (growing faster than linearly) already in the free theory, so we do not have non-trivial tests of our conjecture in such cases (except when the number of fermion fields is very large, as in large $N$ gauge theories, such that the faster-than-linear growth of the free fermion theory sets in only at large charges).

\subsection{The $U(1)$ model in $4-\epsilon$ dimensions}
\label{sec:o2model}

The simplest tests of the conjecture, which we will start from, can be performed in generalizations of the Wilson-Fisher fixed point, which can be analyzed perturbatively in an expansion in small $\epsilon$, where $d=d_0-\epsilon$ for some integer $d_0$. We will show that in all the examples which are believed to correspond to unitary CFTs when $\epsilon=1$, our conjecture holds. However, we should note that strictly speaking the theories with small $\epsilon$ are not unitary CFTs by themselves \cite{Hogervorst:2015akt}, and the extrapolation to $\epsilon=1$ is not under control. So these examples do not provide rigorous tests of the conjecture, but it seems that if we ignore the non-unitarity at non-integer dimensions (which is related to specific operators which vanish as $\epsilon \to 0$), the conjecture does non-trivially hold in all of these cases.
 
The simplest non-trivial CFT of this type that we can consider is the $U(1)=O(2)$ model at the Wilson-Fisher fixed point \cite{Wilson:1971dc}. We follow here the analysis of the anomalous dimension in \cite{Badel:2019oxl}. The Euclidean action is given by
\be
{\cal L} = \partial^{\mu} \phi^{\dagger} \partial_{\mu} {\phi} + \frac{\lambda}{4} \left(\phi^{\dagger} \phi \right)^2 \;.
\ee
In $d=4-\epsilon$ dimensions there is a Wilson-Fisher fixed point (when the mass is fine-tuned to zero) at a coupling 
\be
\frac{\lambda_{\star}}{\left(4\pi\right)^2} = \frac{\epsilon}{5} + \frac{3}{25}\epsilon^2 + {\cal O}\left(\epsilon^3\right) \;.
\ee
The dimension of $\phi^n$ in this model is \cite{Badel:2019oxl}
\be
\Delta\left(\phi^n\right) = \frac{d-2}{2} n + \frac{\epsilon}{10} n \left(n-1\right)
-\frac{\epsilon^2}{100} n \left(2n^2 - 8n+5\right) + {\cal O}\left(\epsilon^3\right)\;.
\label{gphn}
\ee
This gives
\be
\gamma_{n_1,n_2} = \frac{n_1 n_2}{5} \epsilon - \frac{n_1n_2}{50}\left(3n_1 + 3n_2 - 8 \right) \epsilon^2 + {\cal O}\left(\epsilon^3\right)\;,
\label{4du1gn1n2}
\ee
which satisfies (\ref{gccc}) at leading order in $\epsilon$.\footnote{Note that the expression (\ref{gphn}) is derived assuming a small expansion parameter $\epsilon n \ll 1$, and then the first term in (\ref{4du1gn1n2}) always dominates. For large $\epsilon n$ the theory goes over to the large charge regime discussed in the previous section, which is convex. The spectrum remains convex \cite{Gabriel} throughout the continuous interpolation between these two regimes \cite{Badel:2019oxl} (see also \cite{Arias-Tamargo:2019xld,Watanabe:2019pdh}).}

\subsection{The quartic $O(N)$ model in $4-\epsilon$ dimensions}
\label{sec:oNmodel}

More generally, we can consider the $O(N)$ model. Here we will follow the analysis of \cite{Antipin:2020abu,Jack:2021ypd}. The Euclidean Lagrangian is
\be
{\cal L} = \frac12 \partial^{\mu} \phi_i \partial_{\mu} \phi_i + \frac{\lambda}{4!} \left(\phi_i \phi_i \right)^2\;,
\ee
with $i=1,\cdots,N$. In $d=4-\epsilon$ dimensions the fixed point is at
\be
\frac{\lambda_{\star}}{4\pi^2} = \frac{3 \epsilon}{N+8} + \frac{9\left(3N+14\right)\epsilon^2}{\left(N+8\right)^3} + {\cal O}\left(\epsilon^3\right)\;.
\ee
The global group here is non-Abelian, so we need to use the general conjecture (\ref{cccg}); in particular $\phi$ itself is in the fundamental representation of $O(N)$, and we can test the conjecture where $r_0$ is taken to be this representation.

In particular, we can consider the operators $\varphi^n$, where $\varphi \equiv \phi_1  + i \phi_2$; these lie in the $n$'th symmetric product of the fundamental representation of $O(N)$.
Their dimension is given by \cite{Jack:2021ypd}
\be
\Delta\left(\varphi^n\right) = \frac{d-2}{2} n + \frac{\lambda_{\star}n}{3(4\pi^2)}\left(n-1\right) 
- \frac{\left(\lambda_{\star}n\right)^2}{18(4\pi^2)^2}\left( N + 4n - 6 \right) + {\cal O}\left(\left(\lambda_{\star}n\right)^3\right)\;.
\ee
We therefore have
\bea
\gamma_{n_1,n_2} = \frac{2n_1 n_2}{N+8} \epsilon &-& \frac{n_1n_2}{\left(N+8\right)^3} \epsilon^2 \Big( 48n_1 + 48n_2\\ \nn
&&\qquad\qquad\qquad +6 n_1 N + 6 n_2 N - 16 N +N^2 - 132 \Big) + {\cal O}\left(\epsilon^3\right)\;,
\eea
and so (\ref{gccc}) is satisfied at leading order in $\epsilon$.

\subsection{The sextic $U(1)$ and $O(N)$ models in $3-\epsilon$ dimensions}
\label{sec:u1model3e}

For the sextic $U(1)$ model in $d=3-\epsilon$ (where both the $|\phi|^2$ and the $|\phi|^4$ operators are fine-tuned to zero), we will follow the analysis of \cite{Badel:2019khk,Jack:2020wvs}. The Euclidean Lagrangian is
\be
{\cal L} = \partial^{\mu} \phi \partial_{\mu} \bar{\phi} + \left(\frac{\lambda}{3!}\right) \left(\phi \bar{\phi} \right)^3 \;.
\ee 
The fixed point is at
\be
\left(\frac{\lambda_{\star}}{8\pi}\right)^2 = \frac{3}{28} \epsilon + {\cal O}\left( \epsilon^2\right) \;.
\ee
The dimension of $\phi^n$ is given by 
\bea \label{3eu1gn}
\Delta\left(\phi^n\right) &=& \frac{d-2}{2} n + \left(\frac{\lambda_{\star}}{8\pi}\right)^2 \frac{n\left(n-1\right)\left(n-2\right)}{9} \label{u1res}\\
&-& \left(\frac{\lambda_{\star}}{8\pi}\right)^4 \frac{n\left(n-1\right)}{72} \Big[9\pi^2\left(n-2\right)\left(n-3\right) \nn \\
&& \qquad\qquad\qquad\qquad +8n^3 -56n^2 + 272n -456\Big]+ {\cal O}\left(\lambda_{\star}^6\right) \nn \;.
\eea
This gives
\be
\gamma_{n_1,n_2} = \frac{n_1 n_2}{28} \left( n_1 + n_2 - 2\right) \epsilon + {\cal O}\left( \epsilon^2 \right) \;. 
\label{3eu1gn1n2}
\ee
In this case the leading order perturbative correction takes a more complicated form, related to the fact that it arises at two-loop order. In particular we see that for $n_1,n_2 > 1$ we have a positive anomalous dimension at leading order, which is the 2-loop level. But for $n_1=n_2=1$, the two-loop contribution vanishes, and we have to look at the 4-loop contribution. 
Using (\ref{3eu1gn}) to compute $\Delta(2)$, we find
\be
\gamma_{1,1} = 2 \left(\frac{\lambda_{\star}}{8\pi}\right)^4 + {\cal O}(\lambda_{\star}^6) \;,
\label{gam11}
\ee
consistent with (\ref{gccc}).

The $O(N)$ sextic model in $3-\epsilon$ dimensions was also analysed in \cite{Jack:2020wvs}. We can consider the dimension of an operator of total charge $n$ (as in section \ref{sec:oNmodel}), with $r_0$ the fundamental representation of $O(N)$. \footnote{In \cite{Jack:2020wvs} certain contributions to the dimension of $\phi^n$ at four loops were not included, specifically self-energy insertions to the $n$ external legs of the $\phi^n$ vertex. However, since this contribution scales as $n$, it does not modify $\gamma_{n_1,n_2}$.} 
They found
\bea \label{d3Ongn1n2}
 \gamma_{n_1,n_2} = && \left(\frac{\lambda_{\star}}{8\pi}\right)^2\frac{n_1 n_2 \left(n_1 + n_2 - 2 \right)}{3} \\ \nn
& & -\left(\frac{\lambda_{\star}}{8\pi}\right)^4 \frac{n_1 n_2}{72}\Big[ 8 \left( -182 + 123 n_1 - 32 n_1^2 + 5 n_1^3 \right. \\ \nn
& & \left.+ 123 n_2 - 48 n_1 n_2 + 10 n_1^2 n_2 - 
 32 n_2^2 + 10 n_1 n_2^2 + 5 n_2^3 \right) \\ \nn
& & + \left(16 + N\right) \left(11 - 9 n_1 + 2 n_1^2 - 9 n_2 + 3 n_1 n_2 + 2 n_2^2\right) \pi^2 \Big] + {\cal O}(\lambda_{\star}^6)\;.
\eea
%
This is consistent with (\ref{gccc}) for all values of $n_1$ and $n_2$ (including $n_1=n_2=1$).

\subsection{A simple supersymmetric theory}
\label{sec:susy}


Next, consider the supersymmetric theory with four supercharges, a single chiral superfield $\Phi$, and a superpotential 
\be
W = g \Phi^3.
\ee
For $d=3$ this theory is believed to flow to a Wilson-Fisher-like fixed point, while for $d=4$ the theory is IR-free. The only global symmetry of this theory is a $U(1)_R$ symmetry, under which the scalar component $\phi$ of $\Phi$ carries charge $(2/3)$, and its fermionic component $\psi$ carries charge $-(1/3)$. In our conjecture we assumed that the $U(1)$ charges are quantized to be integers, so we should consider three times the $U(1)_R$ charge. Note that already in the free limit of this theory (say in $d=4$), we can see that the lowest-dimension operator with charge $1$, which is $\bar\psi$, has dimension $3/2$, so $\Delta(1)=3/2$, while the lowest-dimension operator with charge $2$ is $\phi$ which has dimension $1$, so $\Delta(2)=1$. This is a simple example where in the phrasing of the conjecture (\ref{ccc}) we cannot take $q_0=1$ but we must take $q_0=2$. However, for $q_0=2$ the conjecture trivially holds in the free limit, since $\Delta(2n)=\Delta(\phi^n)=n$. 

For finite coupling (as in the $d=3$ fixed point), the operator $\Phi$ is a BPS operator so its dimension (and those of $\phi$ and $\psi$) is the same as in the free theory, but the higher $\Phi^n$ are not BPS operators, so they have positive anomalous dimensions (by the BPS bound). As discussed in the previous section, the BPS bound means that our conjecture is always satisfied at leading perturbative order in supersymmetric theories, as long as there is at least one BPS operator charged under the group $G_0$ we are considering (with a charge that is not parameterically large; note that this operator will always carry some R-charge). We do not know how to compute the anomalous dimensions of $\phi^n$ for $d=3$ (beyond the knowledge that $\Delta(\phi^n) > n \Delta(\phi)$). If these can be computed (perhaps by numerical bootstrap methods) they could provide nice tests of our convexity conjecture (e.g. by confirming that $\Delta(\phi^4) \geq 2 \Delta(\phi^2)$).

We can take the Lagrangian of the $d=4$ theory and consider it also in $(4-\epsilon)$ dimensions, keeping four-component fermions. The theory is no longer supersymmetric, but it still has just the $U(1)_R$ symmetry, and it has a weakly coupled fixed point for $g$ that is controllable in the $\epsilon$-expansion, and we can ask if our conjecture it still satisfied. At leading order in the $\epsilon$ expansion the computation of anomalous dimensions (as a function of $g$) is the same as in $d=4$ where the theory is supersymmetric, so we know that $\phi$ has no anomalous dimension (at order $g^2 \propto \epsilon$) while $\phi^2$ has a positive anomalous dimension. On general grounds, as above, the anomalous dimension of $\phi^n$ would be proportional to $\epsilon n (n-1)$ (as in the previous sections where we had one-loop anomalous dimensions), and we know from the $n=2$ case that the coefficient is positive. Thus, our conjecture holds for these theories, but only when we take $q_0=2$.


\subsection{The quartic $O(N)$ model in $d$ dimensions at large $N$}
\label{sec:onmodelgend}

There is an alternative way to compute the anomalous dimensions in the quartic $O(N)$ model of section \ref{sec:oNmodel}, by performing a $1/N$ expansion for general dimension $d$, instead of an $\epsilon$ expansion. The two expansions agree in their overlap. The advantage of this expansion is that it can be used also for the physical case of $d=3$, and thus it gives us a real test of the conjecture (in the large $N$ limit, for $r_0$ the fundamental representation of $O(N)$).
In this case we should consider a general operator of charge $q$, with dimension $\Delta\left(q\right)$. Expanding in $\frac{q}{N}$ for $d=3$ one obtains \cite{Alvarez-Gaume:2019biu,Giombi:2020enj}
\be
\Delta\left(q\right) = N \left[\frac{q}{2N} + \frac{4}{\pi^2}\left( \frac{q}{N}\right)^2 + {\cal O}\left( \left(\frac{q}{N}\right)^3 \right) \right]\;. 
\ee
This gives $\gamma_{n_1,n_2} > 0$, consistent with the conjecture.

It is possible to formally calculate the dimension also in $d=5$ (where the non-trivial fixed point is a UV fixed point rather than an IR fixed point), and this gives \cite{Giombi:2020enj}
\be
\Delta\left(q\right) = N \left[\frac{3q}{2N} - \frac{32}{3\pi^2}\left( \frac{q}{N}\right)^2 + {\cal O}\left( \left(\frac{q}{N}\right)^3 \right) \right]\;. 
\ee 
This leads to $\gamma_{n_1,n_2} < 0$, violating the conjecture. Similarly, one can calculate in $6-\epsilon$ dimensions for $\sigma |\phi|^2$ theories (which are believed to flow to IR fixed points related to the UV fixed points mentioned above) in the $\epsilon$ expansion, and this gives \cite{Arias-Tamargo:2020fow,Antipin:2021jiw}
\be
\gamma_{n_1,n_2} = -264 \epsilon \frac{n_1 n_2}{N^2} + {\cal O}\left( \left(\frac{\epsilon n}{N}\right)^2 \right) \;, 
\ee 
which is again in contradiction with the conjecture. 

However, in $d>4$ (and in particular in $d=5$) it is believed that the CFTs appearing in the computations above are non-unitary, related to the non-boundedness of the scalar potential when we flow to these fixed points from well-defined UV theories
(see, for example, \cite{Giombi:2019upv,Arias-Tamargo:2020fow,Giombi:2020enj}). Thus, the examples of the previous paragraph do not contradict our conjecture, which is only for unitary theories. It is intriguing that if our conjecture is correct, the computations in the previous paragraph can be viewed as providing evidence for the non-unitarity of these CFTs already from perturbative computations, while usually the non-unitarity only shows up through non-perturbative effects.

\subsection{$U(N_c)$ gauge theory in $3$ dimensions with fermions}
\label{sec:qed3mon}

Gauge theories in 3 dimensions with gauge group $U(N_c)$ and $N_f$ fermions in the fundamental representation are expected to flow to an interacting CFT in the infrared for sufficiently large $N_f$. This is true both for QED (with $N_c=1$) and for the non-Abelian case. Such theories have a $SU\left(N_f\right)\times U(1)_{\mathrm{top}}$ global symmetry, where $U(1)_{\mathrm{top}}$ is the symmetry whose current is the Hodge dual of the $U(1)$ field strength. The operators charged under $U(1)_{\mathrm{top}}$ are monopole operators. The spectrum of the monopoles can be studied in a large $N_f$ expansion, and this was done in particular in \cite{Borokhov:2002ib,Pufu:2013vpa,Dyer:2013fja,Chester:2017vdh}, whose results we use\footnote{We thank Shai Chester for suggesting this example.}. 

In perturbation theory, the monopoles are specified by Goddard-Nuyts-Olive (GNO) charges, associated to the Cartan of $U(N_c)$, such that the sum of these charges is the charge under $U(1)_{\mathrm{top}}$. The monopole dimension does depend in general on the specific GNO charges, but at large $N_f$ this dependence drops out, and the dimension is determined just by the $U(1)_{\mathrm{top}}$ charge $q$. 

The monopoles also transform as non-trivial representations of $SU\left(N_f\right)$, with different dimensional representations for different $q$'s. This comes from the quantization of fermion zero modes in the monopole background. To agree with previous sections we rescale the monopole charge by a factor of $2$ compared to the literature, such that it is an integer. In this normalization, the monopoles with $q=1$ transform in some $SU(N_f)$ representation with $N_f/2$ boxes (note that $N_f$ must be even due to the parity anomaly), and all these representations are degenerate at leading order in $1/N_f$. Monopoles with higher $q$'s transform in $q$'th-products of these representations. We can thus take $r_0$ in the formulation of the conjecture (\ref{cccg}) to be some representation with charge $1$ under $U(1)_{\mathrm{top}}$, and some representation (for instance the symmetric one) with $N_f/2$ boxes of $SU(N_f)$. These are the lowest-dimension operators carrying $U(1)_{\mathrm{top}}$ charge; there are higher-dimension operators in other $SU(N_f)$ representations (that one can construct, say, by products of fermions with the simplest monopoles).

The results for the monopole dimension $\Delta_q$ for charge $q$ are given in table \ref{tab:monqed3} taken from \cite{Dyer:2013fja} (rescaling the charges by a factor of 2).
It is simple to check that they are indeed convex at large $N_f$, and so they satisfy the conjecture (\ref{ccc}). 
\begin{table}[!h]
\begin{center}
\begin{tabular}{|c||c|}
\hline
 $q$ & $\Delta_q$ \\
 \hline \hline
 $1$ & $0.265 \,N_f - 0.0383 + {\cal O}(1/N_f)$ \\
  \hline
 $2$ & $0.673  \,N_f  - 0.194 + {\cal O}(1/N_f)$ \\
  \hline
 $3$ & $1.186  \,N_f  - 0.422 +{\cal O}(1/N_f) $ \\
  \hline
 $4$ & $1.786  \,N_f - 0.706 + {\cal O}(1/N_f) $ \\
  \hline
 $5$ & $2.462  \,N_f - 1.04 + {\cal O}(1/N_f) $ \\
 \hline
\end{tabular}
\caption{The first few monopole operator dimensions $\Delta_q$ for monopole charges $q$ in the $U(N_c)$ gauge theory in $3$ dimensions with $N_f$ fermions in the fundamental, taken from \cite{Dyer:2013fja}. \label{tab:monqed3}}
\end{center}
\end{table}

Note that for these monopole operators, as well as for the ones mentioned in the next two subsections and in \cite{Dupuis:2019uhs,Dupuis:2019xdo,Dupuis:2021yej}, the large $q$ expansion works quite well all the way down to $q=1$ (this was first observed in the context of the $\epsilon$ expansion of their dimensions in \cite{Chester:2015wao}). The convexity of the spectrum naturally follows from this behavior. However, note that this statement about the validity of the large $q$ expansion is not true for the other examples we discuss in this paper. Presumably it is related to the fact that the dimensions of the lowest-charge monopole operators are parameterically large (of order $N_f$) in the limit where we can compute them, while the other low-charge operators we discuss have dimensions of order one, and to the fact that the gap to the massive states in the effective theory of the Nambu-Goldstone bosons (expanded around the monopole states) is not small even for $q=1$.

\subsection{$U(N_c)$ Chern-Simons theories in 3 dimensions with fermions}
\label{sec:qed3mon}

The theories above can be easily generalized by adding a Chern-Simons level $k$, as long as $k-N_f/2 \in \mathbb{Z}$. Let us start from the case of $N_c=1$.
Chern-Simons theories with gauge group $U(1)$ in 3-dimensions and Chern-Simons level $k$, with $N_f$ charged fermion fields, flow to an interacting CFT in the IR for large $N_f$ and/or large $k$. The theories have a $U(1)_{\mathrm{top}}$ with monopoles charged under it. The spectrum of such monopoles was studied in \cite{Chester:2017vdh}, which we follow. The simplest monopole operators in this case sit in some $SU(N_f)$ representations that were analyzed there, and we can take $r_0$ to be any one of these representations (since they are all degenerate at large $N_f$). Note that for $k=N_f/2$ the simplest monopole operators are singlets of $SU(N_f)$, so the representation $r_0$ is trivial in $SU(N_f)$.

There is a simple expression for the dimension $\Delta_q$ for large $\kappa \gg 1$, where $\kappa = \frac{k}{N_f}$ (but not necessarily large $q$), which is \cite{Chester:2017vdh}
\be
\Delta_q \simeq \frac23 \left(q \kappa \right)^{\frac32} \;.
\label{largekap}
\ee
This is convex and satisfies (\ref{ccc}). 

For $\kappa$ between 0 and $\frac12$, the spectrum is independent of $\kappa$, at leading order in $N_f$ \cite{Chester:2017vdh}. Therefore, the leading order behaviour is as in table \ref{tab:monqed3}. For $\kappa>\frac12$ which is not very large, the expression for the dimensions is explicit but complicated; however, the spectrum remains convex as a function of $q$ \cite{Chester:2017vdh}. Therefore, in any calculable regime, the theory has a convex spectrum satisfying (\ref{ccc}). 
 
Even though the results above were derived for $U(1)$ gauge group, the same results apply at leading order in $1/N_f$ also for $U(N_c)$. 

\subsection{$U(1)$ Chern-Simons theories in 3 dimensions with scalars}
\label{sec:sqed3mon}

Next, we consider scalar versions of three dimensional QED, involving $U(1)$ gauge theories with $N_f$ charged complex scalars, at Chern-Simons level $k$ \cite{Dyer:2015zha,Chester:2017vdh,Chester:2021drl}. There are two types of such theories which flow to CFTs in the IR (for large enough $N_f$), those with quartic terms, which are commonly known as the $\mathbb{CP}^{N_f-1}$ model, and those where the quartic terms are tuned to zero, which we will refer to as the tri-critical model.

In this case the large $\kappa$ limit again behaves as in (\ref{largekap}), but with a different positive prefactor, $\Delta_q \simeq \frac{1}{\sqrt{2}}( q\kappa)^{3/2}$, and so the spectrum is convex. For small $\kappa$, there is no degeneracy between different values of $\kappa$ as there was in the fermionic case, and the spectrum becomes more convex monotonically for increasing $\kappa$ \cite{Chester:2017vdh}.

In this case the simplest monopoles are $SU(N_f)$ singlets for $\kappa=0$, while otherwise they sit in specific $SU(N_f)$ representations that were analyzed in \cite{Chester:2017vdh}.
For the case of $\kappa=0$ the spectrum for the $\mathbb{CP}^{N_f}$ and tri-critical models at large $N_f$ is given in table \ref{tab:cpntri} as taken from \cite{Dyer:2015zha,Chester}. This is seen to be convex in both cases and so satisfies (\ref{ccc}). One can check \cite{Chester} that the conjecture holds also for non-zero $\kappa$.
\begin{table}[!h]
\begin{center}
\begin{tabular}{|c||c|c|}
\hline
 $q$ & $\mathbb{CP}^{N_f}$  & tri-critical \\
 \hline \hline
 $1$ & 0.125 & 0.097 \\
  \hline
 $2$ & 0.311 & 0.226 \\
  \hline
 $3$ & 0.544 & 0.384 \\
  \hline
 $4$ & 0.816 & 0.567 \\
  \hline
 $5$ & 1.121 & 0.771 \\
 \hline
\end{tabular}
\caption{Table showing the coefficient of the leading term in the $1/N_f$ expansion, namely the large $N_f$ limit of $\Delta(q)/N_f$, for the dimension of the first few monopoles, for the $\mathbb{CP}^{N_f}$ model \cite{Dyer:2013fja} and the tri-critical model \cite{Chester}. 
\label{tab:cpntri}}
\end{center}
\end{table}


\subsection{Banks-Zaks fixed point in 4 dimensions}
\label{sec:bankszaks}

Four-dimensional $SU(N_c)$ gauge theories with $N_f$ flavors of massless charged matter have a perturbative fixed point when the number of flavors is slightly smaller than the value for which the theory becomes IR-free \cite{Caswell:1974gg,Belavin:1974gu,Banks:1981nn}. The coupling constant is parameterically small in the large $N_c$ limit with $N_f$ proportional to $N_c$. The charged matter can be fermions or scalars in the fundamental of $SU(N_c)$.\footnote{In the case of scalars the mass needs to be fine-tuned to zero at each order in perturbation theory.}

When we have scalars, we must also consider $\phi^4$ couplings that are generated by the renormalization group flow, and we need to consider fixed points for these couplings in addition to the gauge coupling. This question was analyzed in \cite{Benini:2019dfy} (see also \cite{Hansen:2017pwe}), where they took $N_s$ scalars and a scalar potential of the form
\be \label{scalarpot}
V(\phi) = {\tilde h} {\rm Tr}(\phi^{\dagger} \phi \phi^{\dagger} \phi) + {\tilde f} ({\rm Tr}(\phi^{\dagger} \phi))^2,
\ee
where $\phi$ is viewed as an $N_c\times N_s$ matrix. They found that for small enough values of $N_s/N_c$, there are perturbative fixed points which are unitary, with real values for the couplings ${\tilde h}$ and ${\tilde f}$ and a positive-definite scalar potential. They found $4$ fixed points of this type, with varying numbers of relevant operators; our analysis below will apply to all of these fixed points. Since $N_s/N_c$ needs to be small, perturbative fixed points of this type require having both fermions and scalars, where the fermions make up most of the contribution to the beta function.

In the case of $N_f$ fermions the global symmetry is $SU\left(N_f\right)_L \times SU\left(N_f\right)_R\times U(1)_B$. We can consider meson operators which are in the bi-fundamental representation of $SU(N_f)_L\times SU(N_f)_R$ and neutral under $U(1)_B$ and their symmetric products, so operators of the form
\be
{\cal O}^{\psi}_{n} = \left(\bar{\psi}\psi\right)^n \;,
\ee
where we can take all the $\psi$'s and all the $\bar{\psi}$'s to have the same flavor index (these operators vanish for $n > 2N_c$, requiring adding derivatives to obtain higher-charge operators, but in the large $N_c$ limit this is not relevant for our considerations).
In the case of $N_s$ scalars the global symmetry is $SU\left(N_s\right) \times U(1)_B$. We can consider again meson operators, of type 
\be
{\cal O}^{\phi}_{n} = \left(\phi^* \phi\right)^n \;,
\ee
where now we choose them to transform in symmetric products of the adjoint representation of $SU(N_s)$. In particular we can choose, for instance, all $\phi$'s to have index $1$ and all $\phi^*$'s to have index $2$, and we will implicitly assume this below.

Consider first the scalar case. We are interested in calculating the difference in dimension between the one-meson and two-meson operators 
\be
\gamma_{1,1}^{\phi} = \Delta\left({\cal O}^{\phi}_{2}\right) - 2 \Delta\left({\cal O}^{\phi}_{1}\right) \;.
\label{ga11def}
\ee 
At leading order in perturbation theory this is sufficient to determine convexity of the spectrum. 
%

\begin{figure}
\centering
 \includegraphics[width=\textwidth]{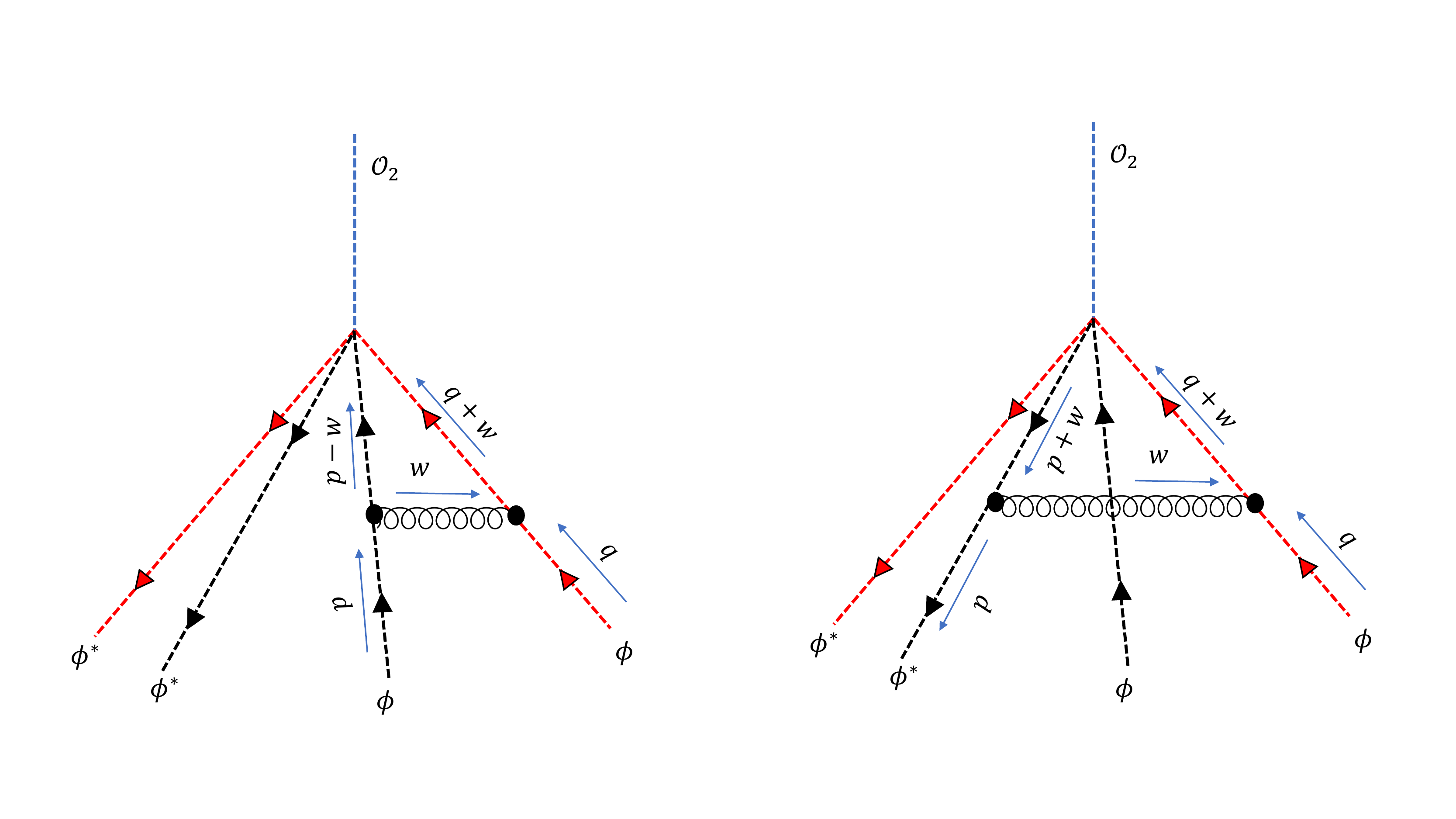}
\caption{The two types of diagrams contributing to the anomalous dimension of the 2-meson operator but not to the 1-meson operator. 
The colors of the lines indicate their color-index-contractions at the ${\cal O}_2$ operator (red with red and black with black).
}
\label{fig:4scalardia}
\end{figure}

There are two types of one-loop contributions to the scalar-meson anomalous dimensions. One contribution comes from gluon exchanges, and it turns out that this contribution exactly vanishes when we consider the combination \eqref{ga11def}.
The reason for this cancellation between the 1-meson and 2-meson anomalous dimensions
can be simply understood as follows. 
To calculate the 2-meson anomalous dimension we consider the correlator\footnote{This correlator is not gauge-invariant, but we can consider it in a specific gauge if we want, and this issue is not important at the leading order we discuss here.}
\be
\left<\phi^*_{j_1}\left(-p_1\right) \phi_{i_1}\left(q_1\right) \phi^*_{j_2}\left(-p_2\right) \phi_{i_2} \left(q_2\right) (\phi^*_{l_1} \phi_{k_1} \phi^*_{l_2} \phi_{k_2}) \left(k\right) \right> \delta^{l_1 k_1} \delta^{l_2 k_2} \;,
\label{2mescor}
\ee
where the first $4$ fields have opposite flavor indices from the ones appearing in the 2-meson operator (the indices written explicitly in \eqref{2mescor} are the color indices).
At 1-loop this receives contributions from 4 types of gluon-exchange diagrams.
First, there are diagrams where a gluon connects a specific external leg to itself. These are wavefunction regularizations, that cancel when computing differences such as (\ref{ga11def}). Next, there are those diagrams where the gluon is exchanged between scalars in the same meson, so say between $\phi^*_{l_1}$ and $\phi_{k_1}$. Those will also cancel with the 1-meson contributions when we take the difference in dimensions (\ref{ga11def}).
The other two types of diagrams are where the gluon is exchanged between the mesons, so say between $\phi^*_{l_1}$ and $\phi^*_{l_2}$. In one set of diagrams the gluon is exchanged between two $\phi$'s (or two $\phi^*$'s), and in the other it is exchanged between a $\phi$ and a $\phi^*$. The two possibilities are shown in figure \ref{fig:4scalardia}.

When computing the 2-meson diagrams in figure \ref{fig:4scalardia},
it is easy to see that there is a sign difference between (the logarithmic divergence of) the two diagrams in figure \ref{fig:4scalardia} (while their absolute value is equal). This is essentially because in the diagram on the left the two vertices contribute $(2q+w)_{\mu}(2p-w)_{\nu}$, while in the diagram on the right the two vertices contribute $(2q+w)_{\mu}(2p+w)_{\nu}$, while all other factors are the same (and only the $w^2$ term from the vertices contributes to the logarithmic divergence). So, diagrams where the gluon is exchanged between $\phi^{*}$ and $\phi$ contribute to the anomalous dimension with opposite sign to those where the gluon is exchanged between $\phi$ and $\phi$ (or $\phi^*$ and $\phi^*$). Therefore, the diagrams cancel between themselves, leading to an overall vanishing contribution. It is easy to see that the same cancellation occurs in any computation of $\gamma_{n_1,n_2}$.

Thus, the only one-loop contribution to $\gamma_{n_1,n_2}$ comes from the $\phi^4$ couplings \eqref{scalarpot}. Note that like the previous contributions, this contribution is independent of the number of fermions. It is not difficult to compute the contribution of the couplings above to \eqref{ga11def}, and we find that it is proportional to $(\tilde{f}+\tilde{h})$, with a positive coefficient. It was already noted in \cite{Benini:2019dfy} that this combination is positive in all the fixed points, since this is related also to the positivity of the scalar potential. Thus, this example is consistent with our conjecture.

We can also consider other gauge+scalar fixed points where one can compute multi-meson dimensions perturbatively, and we will similarly find that the gluon exchange amplitudes cancel at one-loop order. In particular this applies also to the $3d$ fixed points for $SU(N_c)$ (or $U(N_c)$) theories with $N_f$ scalars in the fundamental representation, where perturbation theory is valid for large $N_f$. In this case (where some anomalous dimension computations were performed for fermionic theories in \cite{Chester:2016ref}) there are no $\phi^4$ couplings (which are fine-tuned to zero), so the one-loop multi-meson anomalous dimensions cancel, and we need to go to two-loop order to check our conjecture.

In the case of fermions, the situation is more subtle,
since we have several different meson operators of different spins in the same bi-fundamental flavor representation (and with the same classical dimension), and even more operators (including several operators of the same spin) when we look at two-meson operators.
For instance we can consider
\be
\bar{\psi} \gamma^{\mu} \psi \bar{\psi} \gamma_{\mu} \psi \;, \;\; \bar{\psi} \gamma^{\mu\nu} \psi \bar{\psi} \gamma_{\mu\nu} \psi \;, ...
\ee
which will mix with the two-scalar-meson operators. Testing the conjecture then requires keeping track, at each level $n$, of the lightest operator, while carefully taking the mixing into account. This is a complicated task, even at the 1-loop level, and so we leave it for future work. This issue arises both for the $4d$ Banks-Zaks fixed point, and for $3d$ fixed points (where the fermion has only two spin indices rather than four, but still there are several different meson and multi-meson operators with the same classical dimension and the same flavor representations, complicating the computation).


\section{Summary}
\label{sec:summary}

In this paper we pointed out that the natural formulation of the Weak Gravity Conjecture in AdS space is in terms of the self-binding energy of a charged state. This has a simple CFT dual, which is that there should exist a charged operator such that its products with itself have a non-negative anomalous dimension. This formulation generalizes naturally to all CFTs, even without weakly-curved gravitational duals, in the form of the conjectures that we presented in the introduction.

In this paper we performed some simple preliminary tests of the conjecture, and we found that it is always satisfied. It would be interesting to perform more tests of the conjecture, in various additional CFTs where operator dimensions can be computed (for instance, one could look at those fixed points found in \cite{Antipin:2021akb} which are unitary). It would also be interesting to study various generalizations of the conjecture, for instance to non-relativistic CFTs, or to operators living on boundaries or defects in CFTs.

Assuming that the conjecture is correct, it would be very interesting to try to derive it. One option is to use conformal bootstrap methods to show that the spectrum must always be convex. Since we presented an example where $q_0$ needs to be larger than one for the conjecture to hold, it is not completely clear how to do this; but it is possible that convexity always holds for scalar charged operators, and this seems like the simplest case to try to prove by bootstrap methods. Note that just by conformal bootstrap methods one cannot disprove the conjecture, since generalized free fields trivially obey the conjecture and are consistent with the bootstrap conditions (it may be possible to numerically find solutions to the bootstrap equations where the conjecture would not hold, but one would still need to prove that these correspond to full-fledged CFTs). Note also that it seems challenging to derive the conjecture perturbatively, since we have examples above (like the large $N$ $O(N)$ model in $d=5$) where the conjecture does not hold even though the CFT is unitary in perturbation theory (and non-unitarity only appears non-perturbatively in the $1/N$ expansion; in particular, numerical bootstrap computations at current levels of precision lead to an apparently unitary solution to the bootstrap equations for the $d=5$ large $N$ $O(N)$ model \cite{Chester:2014gqa,Li:2016wdp}, even though the full CFT is believed to be non-unitary.).

\vskip 20pt
\noindent {\bf Acknowledgements:} We thank N. Arkani-Hamed, S. Chester, G. Cuomo, J. Gracey, B. Heidenreich, S. Hellerman, I. Jack, Z. Komargodski, A. Kovner, D. Kutasov, M. Lublinsky, J. Penedones, T. Rudelius, T. Ryttov and M. Watanabe for useful discussions. The work of OA was supported in part by an Israel Science Foundation center for excellence grant (grant number 2289/18), by grant no. 2018068 from the United States-Israel Binational Science Foundation (BSF), and by the Minerva foundation with funding from the Federal German Ministry for Education and Research. The work of EP was supported by the Israel Science Foundation (grant No. 741/20). OA is the Samuel Sebba Professorial Chair of Pure and Applied Physics.

\bibliographystyle{ssg}
\bibliography{susyswamp.bib}  
\end{document}